# Unlocking Extra Value from Grid Batteries Using Advanced Models

Jorn M. Reniers[1,2,3], Grietus Mulder[2,3], David A. Howey[1,4]*

[1] Department of Engineering Science, University of Oxford, Oxford OX1 3PJ, United Kingdom

[2] EnergyVille, Genk 3600, Belgium

[3] VITO, Mol 2400, Belgium

[4] Faraday Institution, Harwell Campus Didcot OX11 0RA, United Kingdom

* correspondence: david.howey@eng.ox.ac.uk

Abstract

Lithium-ion batteries are increasingly being deployed in liberalised electricity systems, where their use is driven by economic optimisation in a specific market context. However, battery degradation depends strongly on operational profile, and this is particularly variable in energy trading applications. Here, we present results from a year-long experiment where pairs of batteries were cycled with profiles calculated by solving an economic optimisation problem for wholesale energy trading, including a physically-motivated degradation model as a constraint. The results confirm the conclusions of previous simulations and show that this approach can increase revenue by 20% whilst simultaneously decreasing degradation by 30% compared to existing methods. Analysis of the data shows that conventional approaches cannot increase the number of cycles a battery can manage over its lifetime, but the physics-based approach increases the lifetime both in terms of years and number of cycles, as well as the revenue per year, increasing the possible lifetime revenue by 70%. Finally, the results demonstrate the economic impact of model inaccuracies, showing that the physics-based model can reduce the discrepancy in the overall business case from 170% to 13%. There is potential to unlock significant extra performance using control engineering incorporating physical models of battery ageing.

## 1. Introduction

The amount of renewable energy being generated is increasing and the cost of lithium-ion batteries for grid applications decreasing, accelerating uptake [1]–[3] As a result, merchant energy trading in wholesale and balancing markets using batteries is growing in attractiveness to investors. The size of the trading market is substantially larger than the market for ancillary services – for example, in Germany the total primary frequency control requirement is 650 MW peak power, but there is 40 GWh energy storage capacity in pumped hydro available for trading [4].

Grid battery assets are long-term investments with relatively large capital costs. A crucial part of the business case is the asset lifetime, and this is heavily dependent on usage. However, battery degradation is complex, involving interacting chemical, electrical, mechanical and thermal effects [5]–[7]. Therefore, the cost of taking a certain charge or discharge control action is not constant but is instead a function of the present state and history of a battery.

Various previous authors have simulated the economic opportunities available from grid batteries, and the most common approach is to solve an optimisation problem to maximise profit, assuming known energy prices and an empirical model for degradation [8]–[11]. Such a model would typically fix the degradation cost per time step as a function of the charge throughput and normalised power. For example Fortenbacher et al. [12] demonstrated that a 30% decrease in losses and a doubling of battery lifetime could be achieved. One weakness of these approaches is that their degradation models are simple curve fits from laboratory test data.

A more general approach exploits knowledge of the underlying physics of degradation. Weißhar et al. [13] and Patsios et al. [14] suggested, by using physics-based models in simulation case studies, that degradation could be decreased by over 50%. Our earlier simulation study showed that use of more physically realistic models could increase the revenue from a battery trading on the day-ahead power market by 40% whilst decreasing degradation by 30% [15].

In summary, it has been suggested that including realistic degradation models within a battery control strategy strongly improves revenue, profit and lifetime. However, no previous work has demonstrated these results experimentally, therefore it is unclear whether the additional complexity required by higher fidelity models is justified. The innovation in this work is that lithium-ion cells were cycled with profiles optimised for energy trading to experimentally validate degradation models of differing complexities. The experiments confirm that profiles associated with physics-based models can simultaneously increase revenue and decrease degradation over time, by respectively 20% and 30%, and decrease degradation per cycle by 70%.

## 2. Methods

**2.1 Modelling and optimisation**

In this work, cells were cycled with profiles for trading energy on the wholesale arbitrage market. The profiles were optimised to maximise revenue and minimise degradation. The optimisation framework is detailed in [15] and shown in Fig. 1. A summary is presented in this section, highlighting the assumptions which explain some of the results observed in the experimental validation, and the key equations are given in Appendix A.

We compared two different approaches for maximising net financial returns using a battery for energy trading, subject to physical constraints: (1) The conventional approach is based on a linear battery model parametrised from the cell's data sheet; this was then used in a linear optimisation algorithm, with historical energy price data, to calculate a charge/discharge profile. (2) The physics-based approach on the other hand first characterised the degradation of the cells experimentally. From a number of available physics-based models, the ones which best describe the observed behaviour were selected and fitted to the data. Then a nonlinear optimisation algorithm was used to find an 'optimal' usage profile, again using historical energy price data, with the initial guess taken from the conventional approach.

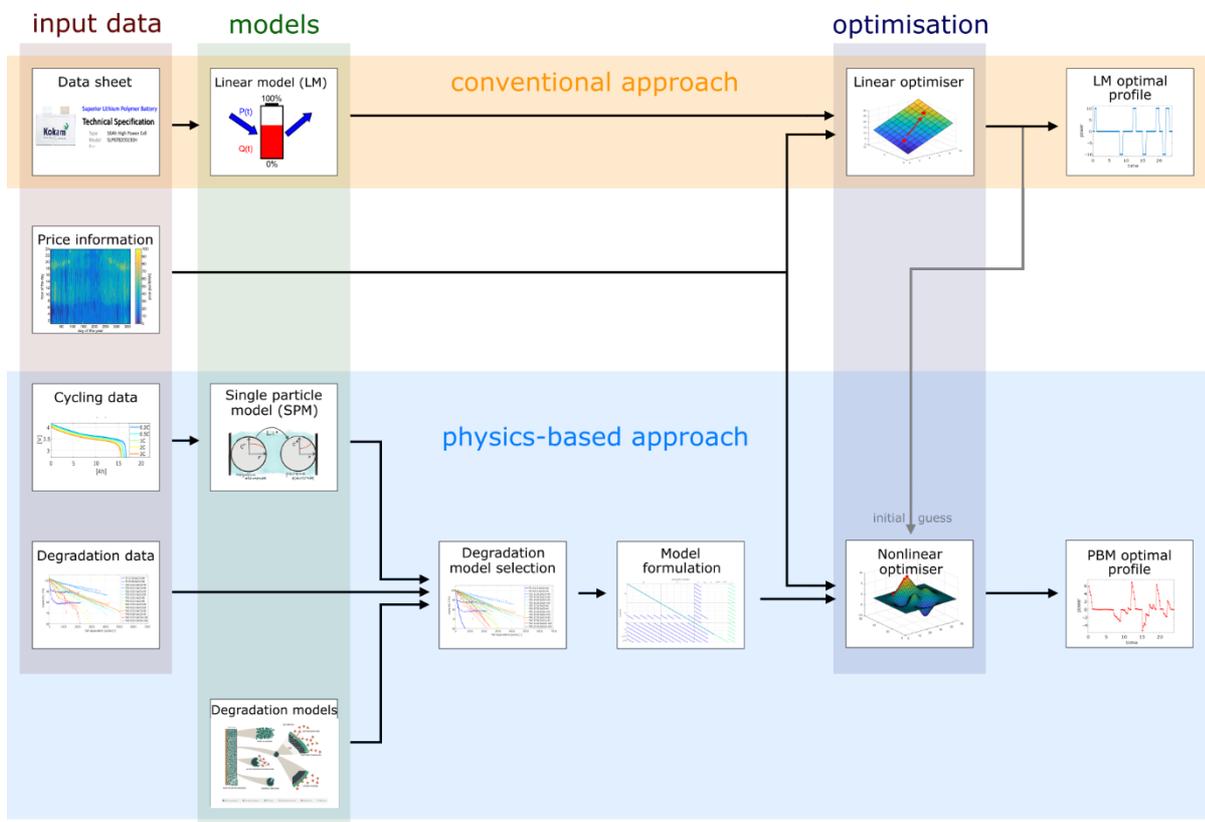

Figure 1: Overview of the two approaches used to obtain the usage profiles for battery arbitrage

Degradation is accounted for by applying a linear correlation that penalises the charge throughput and maximum power, without state-of-charge dependency or calendar ageing. Such a model is often used by non-battery experts, e.g. [16]–[20], because it can be parametrised purely based on information from the limited data sheet of the cell. Occasionally, a simple calendar ageing model where degradation is only a function of time is added as well, e.g. [21], but this is not included here. The physics-based model (PBM) used in [15] is a single particle model (SPM), which is a simplified electrochemical battery model. It is relatively accurate, in terms of voltage prediction, for currents up to about 1C [22]. It simulates lithium diffusion in solid particles using Fick's law, and lithium intercalation using Butler-Volmer kinetics. The single particle model was extended with a bulk thermal model and an Arrhenius relation for temperature dependency of rate- and diffusion constants.

There are various degradation mechanisms which reduce the performance of a Li-ion battery. Different models exist for each mechanism and an overview is given in [23]. The degradation model used in [15] is a diffusion and kinetically limited model for solid electrolyte interphase (SEI) layer growth, as proposed by Christensen et al. [24]. This model was chosen because it fits the degradation data of the cell used in this work reasonably well as shown in [15] and it describes a degradation mechanism which is often said to be the main determinant of battery ageing [25].

The parameters of the linear model, and the single particle model with the SEI growth equation were fitted to a large degradation set for an NMC/graphite cell [26]. The parametrisation of the physics-based model was done using an exhaustive search, details are given in [27]. A comparison between the models and data is given in [15].

The grid energy storage application considered in this work is energy arbitrage. The electricity price $\lambda_{el}$ was assumed to be known perfectly, and a dataset from the day-ahead market in Belgium in 2014 was used. Fig. 2 gives examples of this for different seasons (the full dataset is presented in supplementary material, Fig. S1).

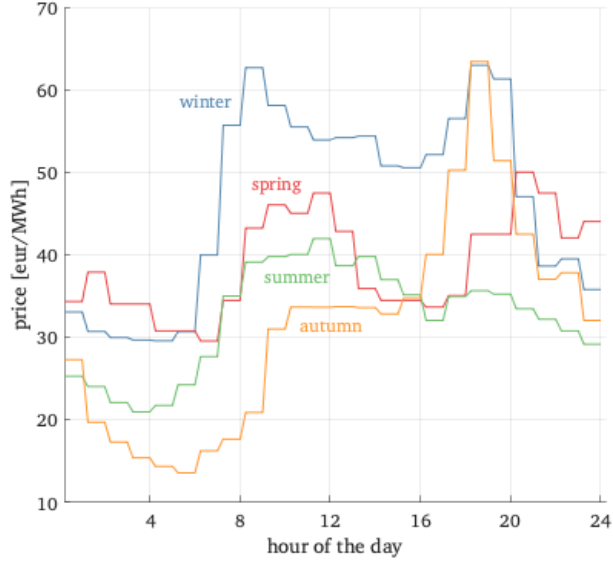

Figure 2: Examples of electricity prices on the day-ahead market in Belgium in 2014.

Although revenue is made by trading energy (i.e. charging at low prices and discharging at higher prices), the associated degradation from cycling leads to an additional cost, $\lambda_{\text{deg}}$, which was assumed here to be fixed.

Both objectives (revenue maximisation and degradation minimisation) can be combined into a single objective function with weighting factor $\theta$:

$$\pi = \int_{t_0}^{t_n} \left( \theta P(t) \lambda_{\text{el}}(t) - (\theta - 1) \frac{dC}{dt} \lambda_{\text{deg}} \right) dt. \tag{1}$$

If $\theta = 1$, the objective is to maximise revenue while ignoring degradation, and if $\theta = 0.5$ then equal weight is given to maximising revenue and minimising degradation.

The solutions to the optimisation problems are four power or current profiles for a full year of operation, two for each combination of the two battery models (linear model or physics-based model), one ignoring degradation and maximising revenue, and the other maximising net profit, i.e. co-optimising for revenue and degradation. Fig. 3 shows examples of these profiles for a single day of operation in winter (the full dataset is presented in Supplementary Material, Fig. S2). Profiles have daily patterns, as shown here, and also seasonal variability (see Fig. S2). These patterns are the result of similar patterns in the price of the wholesale market.

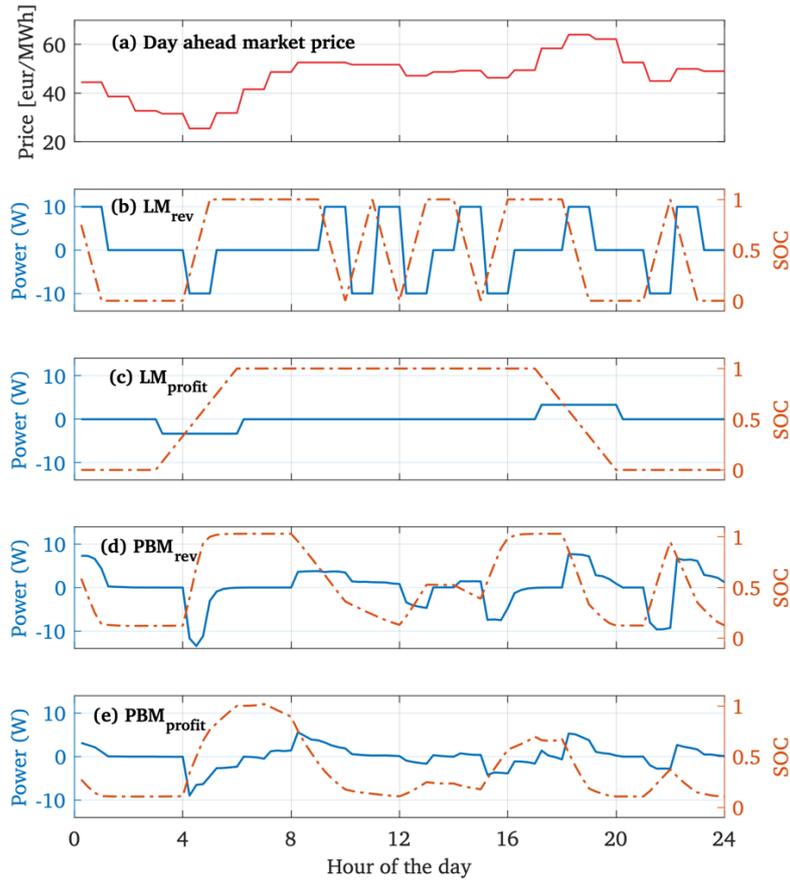

Figure 3: Example usage profiles resulting from the four optimal solutions against price information (a), using: (b) linear model maximising revenue; (c) linear model maximising profit; (d) physics-based model maximising revenue; (e) physics-based model maximising profit.

The profiles optimised according to the linear model almost always fully charge or discharge the battery at constant power, while the profiles resulting from the physics-based model follow more complex cycling behaviour, often at lower power levels and correspondingly longer charging times. For example, to reduce calendar ageing, the physics-based model approach often only partially (dis)charges the cell, unless the price spread is sufficient to overcome the associated increased degradation cost. When accounting for degradation, the optimal profile according to both the linear and the physics-based model uses the battery at lower power levels compared to revenue-maximising optimisations ignoring degradation.

## 2.2 Experimental setup

Lithium-ion cells were cycled using a battery tester (PEC SBT8050) according to the optimised profiles. While cycling, the cells were located on a table in a room which was temperature-controlled to 25 °C ± 0.5 °C, but there was no fan to cool the cells as would be the case in a thermal chamber. The 16 Ah pouch cells chosen (Kokam SLPB78205130H) had a nickel manganese cobalt cathode and graphite anode (NMC / C). These cells are appropriate for stationary grid energy storage applications, having long lifetime and high efficiency. All profiles were rescaled appropriately to fit the 16 Ah nominal capacity instead of the 2.7 Ah (10 Wh) cell used in the modelling. The results given in the rest of this work are scaled down to the 2.7 Ah (10 Wh) cell size to enable accurate comparison.

Due to test channel constraints, only six cells could be tested concurrently, therefore only three usage profiles could be experimentally validated since every profile was tested on two cells. Because the revenue-maximising profiles from both the physic-based model and linear model produced very similar outcomes, only the linear model revenue-maximising profile was tested.

The PEC battery tester was programmed to follow the current profiles while satisfying the voltage limits of the cells. If a voltage limit was reached during a step in the profile, the voltage was kept constant for the remaining time in that step, both during charge and discharge. This ensured that each cell delivered its maximum power without violating safety limits. Voltage limits were only occasionally reached for the physics-based model profile. However, the cells tested with the linear model optimised profiles often reached their voltage limits, because the profiles instructed the cells to undertake a high-power charge/discharge up to the maximum/minimum SoC, while the maximum/minimum voltage will in that case be reached before the SoC limit.

In the revenue-maximising LM and the PBM tests, the voltage limits were set to the full range of the cell, 2.7 V to 4.2 V, while in the profit-maximising linear model tests cells were forced to operate in a smaller voltage range of 3.42 V to 4.08 V, corresponding to 10% and 90% SoC respectively. This reduced state of charge window reflects the regular practice to limit the usable capacity to 80% of the theoretical full capacity to account for safety limits and model uncertainty [28], and to avoid excessive degradation [29], [30]. To analyse whether or not reducing the usable capacity is a valuable strategy to reduce degradation, these limits were not enforced in the revenue maximising case. The physics-based model accounts for the state-of-charge dependency of degradation and therefore usable capacity limits do not need to be enforced externally in that case.

Experiments were interrupted monthly and a separate check-up test undertaken to measure the remaining capacity directly. This consisted of fully charging and discharging the cells three times (CCCV at 1C, to a limit current of 0.01C, with a 1 h rest between charging and discharging). There were a few interruptions in the experiments, due to recalibration, power cuts, and software failures. Overall, 347 days of data were collected.

## 3. Results

### 3.1 Experimental validation of degradation and revenue

During experiments the power and voltage were recorded every 15 minutes. This dataset is publicly available as outlined in the supplementary materials section. The 2D-histograms of mean measured power and voltage across cell pairs following the same charge/discharge profile, grouped per 0.1 V and 1 W, are shown in Fig. 4, with different colour scales used to indicate when the cells were resting compared to when they were cycling. The revenue-maximising linear model cells, Fig. 4(a), spent about 40% of the total time cycling, and this was always at a 1 C power (10 W) because the market price is constant for one hour. When resting, the cells were always fully charged or discharged, although relaxation effects meant that the cells were resting slightly below the maximum and above the minimum voltage.

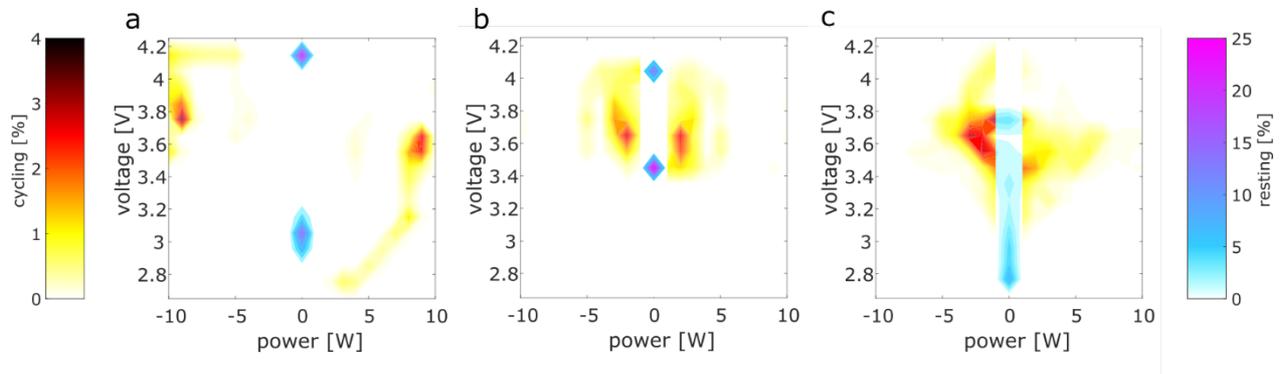

Figure 4: Percentage of time spent by both cells following each profile, as a function of power and voltage. Resting is indicated on a different colour map due to the different order of magnitude. Negative power levels indicate charging. **a** Revenue-maximising linear model; **b** Profit-maximising linear model; **c** Profit-maximising physics-based model.

The profit-maximising linear model cells, Fig. 4(b), were constrained by narrower voltage limits corresponding to safety limits on the SoC to avoid over(dis)charging and the resulting increased degradation. Therefore, the cells rested at 3.42 V or 4.08 V (corresponding to 10% and 90% SoC). The cells cycled for about 40% of the total time, but the average power level was below 0.5 C.

The physics-based model cells, Fig. 4(c), avoid the high-voltage operating region during both cycling and resting. On the few occasions when the price differential was large enough to justify the additional degradation, the cells were fully charged. More often, the cells were only partially charged, with the target voltage level determined by the trade-off between degradation and revenue due to price spread. Cells were almost always fully discharged to a low voltage to reduce degradation during resting. The discharge often happened at higher power compared to the charge, explaining why the cells spent less time discharging compared to charging. In total, the cell was cycled for about 60% of the time, which is higher than for the linear model cells, because the power levels of the physics-based model cells are on average lower than those of the linear model cells, around 0.33 C, and lower when the SoC is higher.

Fig. 5 shows the measured remaining capacity for all cells tested (within each pair, for profit-maximising linear model and physics-based model scenarios, results from individual cells are very similar and overlap). Capacity checks were done monthly, and Fig. 5a shows the degradation every month. While cells were cycling, the total charge throughput was recorded by the battery cycler, such that at each check-up, the number of full equivalent cycles (FEC) since the previous check-up could be calculated ($FEC = \int \frac{|I|}{2Q} dt$ where $I$ is current and $Q$ is nominal capacity). The degradation as function of this battery use is shown in Fig. 5b (again, within each pair, for profit-maximising linear model and physics-based model scenarios, results from individual cells are very similar and overlap). The aggressive use profile associated with the revenue-maximising linear model resulted in severe degradation, losing 14% of capacity within a year.

Conversely, co-optimising for degradation using the linear model reduced degradation to 2.5% in a year, because the cells avoided the extreme SoC regions and used lower power levels. However, when expressed as a function of full equivalent cycles, Fig. 5(b), the linear model profit-maximising cells degraded at the same rate as the revenue-maximising cells. The profit-maximising physics-based model cells reduced their degradation both as a function of time and as a function of full equivalent cycles. After one year, the degradation was reduced by 30% (from 2.5% to 1.7%) while after 386 FEC, it was reduced by about 75% (from 2.5% to 0.6 %) compared to the profit-maximising linear model cells. This is possible because both the calendar and cycle ageing were reduced – resting at lower voltages reduces calendar ageing, while cycle ageing is reduced by using variable currents that depend on the SoC, temperature and energy price spread.

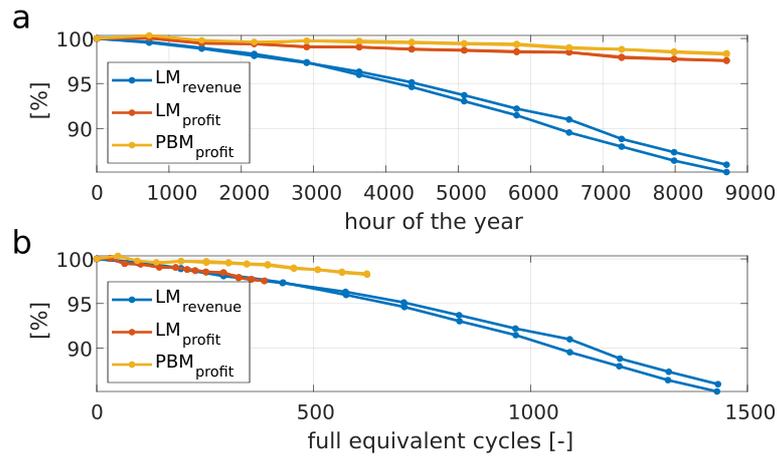

Figure 5: Normalised capacity for each cell, measured in check-up tests: **a** versus time; **b** versus full equivalent cycles.

Fig. 6 shows the revenue that each cell could have made by trading power in the market assuming perfect price information. Due to their large number of cycles, the revenue-maximising linear model cells reached the highest revenue. Over the same time period, the physics-based model cells' revenue was 6% lower, but still 17% higher than the revenue from the profit-maximising linear model cells. In this case the revenue per cycle increases the fewer times a cell is cycled because it will only cycle during periods of higher price spreads. The physics-based model cells are able to reduce their degradation, making cycling at lower price spreads profitable. Therefore they have a lower revenue per cycle compared to the profit-maximising linear model cells, but are able to cycle more over a given time period.

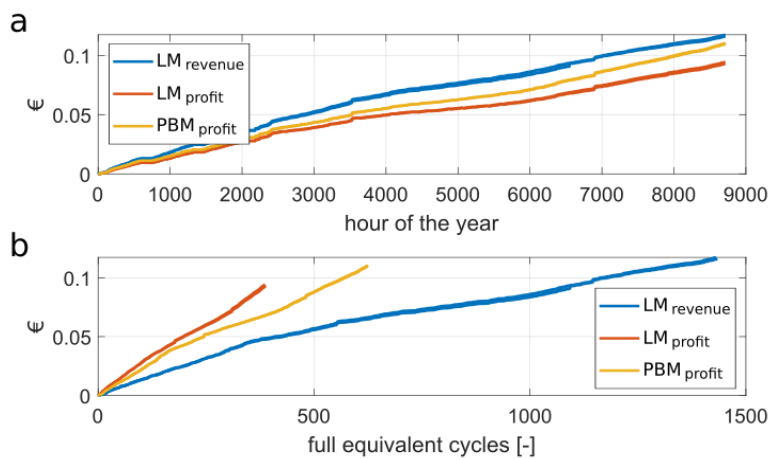

Figure 6: Revenue made by each cell on the wholesale market: **a** versus time; **b** versus full equivalent cycles.

The profit-maximising linear model cells lost 2.46% of their capacity in one year of operation. Linearly extrapolating this to the full lifetime, defined as the period until the cells have lost 20% capacity, the lifetime would be about 8.2 years, and each cell would have a total revenue of 0.76 €, or 76.2 €/kWh. On the other hand, cells from the physics-based SPM profiles lost 1.71% capacity and would be operational for about 11.7 years, reaching a total revenue of 129 €/kWh, a 70% increase compared to the profit-maximising linear model.

## 3.2 Comparison of degradation modelling approaches

The rate of degradation of a lithium-ion battery is not constant but depends strongly on usage. If degradation is ignored in economic optimisation for arbitrage, and consequently only revenue is maximised, the resulting usage profile will be aggressive. It will include multiple high-power cycles per day, and often will rest the battery at high SoC, leading to severe shortening of life. A commonly applied heuristic solution to reduce degradation [13], [14], [20], [29]–[31] is to include static limits for the power and SoC, and to add a fixed cost per cycle within the objective function of the economic optimisation. The static limits avoid aggressive cycling and resting periods at high SoC, and the cycling cost ensures that the battery is only used if the revenue exceeds some threshold price spread. In our experiments, this approach (the profit-maximising linear model) decreased capacity loss from 14% to 2.5% over one year of operation, while reducing the attainable revenue by about 20%.

However, this heuristic approach works by simply reducing the charge throughput of the battery, reducing both revenue and degradation over time. The battery is only used if the potential revenue is large enough, but the longer resting periods between subsequent cycles still degrade the cell. Therefore, the degradation per cycle does not change. This suggests that although the battery could therefore operate usefully over more years, it will still achieve the same total number of cycles in its lifetime.

Fig. 7 illustrates this effect over one day of cycling. Using the Mat4Bat dataset [26], we linearly interpolated the approximate degradation rate as a function of time and cycles; for instance, cells lose $4.2 \times 10^{-4}$ % capacity per hour for resting at 100% SoC, and $6.7 \times 10^{-3}$ % per cycle between 0% and 100% SoC. During an arbitrarily chosen day, the profit-maximising linear model cycled once, corresponding to about 0.8 full equivalent cycles. The revenue-maximising linear model reached 0.8 FEC ten hours earlier as indicated by the dashed magenta arrow. Fig 7(b) shows the degradation at the corresponding points. Summarising, both cells experienced about 0.08% degradation after 0.8 FEC, but after different amounts of time. Therefore, the profit-maximising linear model cannot increase the number of cycles the battery will be able to complete over its lifetime.

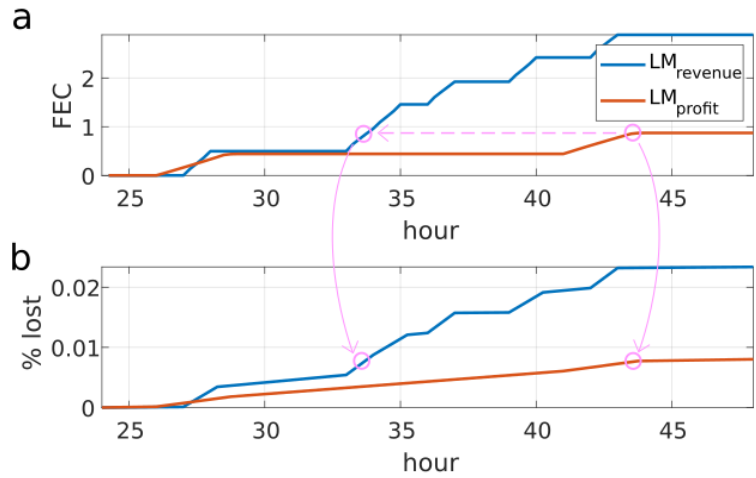

Figure 7: Analysis of the linear model cells' degradation during a single day of testing
**a** battery use expressed in full equivalent cycles over time; **b** estimated degradation.

The physics-based modelling approach explored in this work resulted in a more complex usage pattern, where the cell was often only partially (dis)charged, at varying power levels. Therefore, the degradation over one year of operation was reduced to 1.7%, about a third less than the heuristic approach of the profit-maximising linear model, and 90% less than the linear model ignoring degradation. When the price spreads were large, the full capacity of the cell could be accessed because the physical model accounts for diffusion, reaction and resistive overpotentials, and therefore a (dis)charging profile that maximised the usable capacity of the cell was produced. As a result, the revenue is 17% higher than in the heuristic case, which can never access the full capacity due to the static limits on the SoC.

Significantly, the optimisation using the physics-based model was also able to reduce the *degradation per cycle* compared to the linear model approaches. The reason is that the latter will typically require the battery to undertake constant power (dis)charge cycles, while usage patterns dictated by the physics-based model result in cycles which reduce damage by using a variable power and by resting at lower SoCs. As a result, the number of cycles and arbitrage revenue/profit achievable over the lifetime is increased, Fig. 8. The gradient in this figure shows the revenue per % degradation, a simple and effective metric for comparing the results. Therefore, the PBM approach is superior compared to the heuristic approach, which can only extend the time over which the battery might operate, but not the number of cycles. This also indicates how both goals of the objective function, increasing revenue and decreasing degradation, can be improved at the same time – it is not simply a trade-off between them.

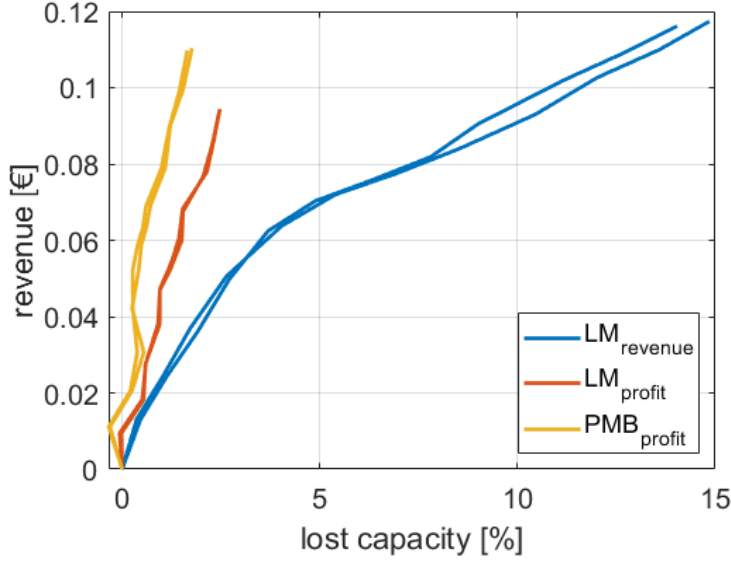

Figure 8: Revenue plotted as a function of lost capacity comparing the linear models maximising revenue and profit, and the physics based model maximising profit.

### 3.3 Comparison of simulated versus measured performance

Table 1 compares the simulated outcomes for revenue and degradation against the experimental measurements. The revenue-maximising linear model significantly overestimates revenue because it does not account for battery voltage limits, which are reached sooner than SoC limits. The profit-maximising linear model introduced static limits on SoC, and is therefore more accurate in revenue predictions, but this comes at the cost of a significantly reduced revenue because the volume of energy traded is lower. The physics-based model directly accounts for the cell voltage, including internal losses, and can therefore also accurately predict the revenue, which is larger than the profit-maximising linear model. Similarly, the degradation prediction versus experiment using the physics-based model is more accurate, with an error of 0.3 %-pts on the final value. The linear models had errors of 2.5 %-pts and 1 %-pts respectively for the profit- and revenue-maximising cases.

Table 1: Comparison between simulations and experimental results after one year of operation; 'LM$_{revenue}$' refers to the revenue-maximising linear model, 'LM$_{profit}$' refers to the heuristic approach using a linear model and static limits, co-optimising revenue and degradation, and 'PBM$_{profit}$' refers to the physics-based model co-optimising revenue and degradation.

|  | Revenue [€] | | Degradation [% of initial capacity lost] | |
| --- | --- | --- | --- | --- |
|  | Simulation | Experiment | Simulation | Experiment |
| LM$_{revenue}$ | 0.194 | 0.117 | 11.87 | 14.44 |
| LM$_{profit}$ | 0.096 | 0.094 | 3.44 | 2.46 |
| PBM$_{profit}$ | 0.115 | 0.110 | 2.03 | 1.71 |

Using the degradation cost of 330 €/kWh from the optimisation, the cost associated with the 2.46% degradation of the profit-maximising linear model cells in one year of operation is about 0.081 €. The revenue was 0.094 €, resulting in a net profit of 0.013 €, or 1.28 €/kWh, while the optimisation outcome predicted a net loss of 1.74 €/kWh. The discrepancy of the predicted versus the measured outcome is 170%, which is explained by the inaccurate degradation prediction. On the other hand, the physics-based SPM is much more accurate, predicting a yearly revenue of 4.78 € while the measured value was 5.39 €, an error of only 13%.

## 4. Discussion
### 4.1 Comparison with literature
We now further discuss these results from two perspectives and contrast them with work by others, firstly in terms of the lifetime prediction accuracy of different modelling approaches, then secondly from an economic perspective.

Many simulation studies have been performed by others to analyse the impact of degradation in battery grid storage applications. Some studies do not include any degradation data and simply propose a model, some start from an initial batch of data to parametrise a model, and then run an open-loop optimisation, and some take model parameters from literature. Table 2 gives an overview of the results from literature, comparing the (simulated) lifetimes of batteries optimised for various grid applications. Lifetime is defined as the period until the remaining capacity is 80% of the initial value, and degradation was extrapolated linearly in cases where only a fraction of the lifetime was simulated. When papers proposed multiple optimal profiles according to different control strategies, the range is given in Table 2. The large variability between studies is due to the varying applications and assumptions considered and the simple degradation models used.

Table 2: Simulated lifetimes of batteries cycled according to optimal profiles for different grid applications. Third column indicates availability of degradation data used for model parameterisation ('other' means a different study is referenced in this regard)

|  | Application | Degradation data | lifetime [years] | Cycle life [cycles] |
|---|---|---|---|---|
| Cao et al. 2020 [32] | Frequency control | no | Up to 1.5 | 290 to 1,020 |
| Corengia et al. 2018 [33] | Time-of-use tariff | no | 4 to 20 |  |
| Fleer et al. 2016 [34] | Frequency control | no | 15 | 3,945 to 7,582 |
| Fortenbacher et al. 2017 [12] | Peak shaving | other | 0.3 to 4.4 | 166 to 2,816 |
| Goebel et al. 2017 [35] | Secondary reserves | other | 5 to 7 | unknown |
| Patsios et al. 2016 [14] | Peak shaving | no | 0.3 to 0.6 | unknown |
| Perez et al. 2016 [28] | Multi-service | other | 6 to 20 | unknown |
| Smith et al. 2017 [36] | Self-consumption | yes | 2.5 to 4 | unknown |
| Stroe et al. 2016 [37] | Frequency control | yes | 10.2 | unknown |
| Weisshar et al. 2017 [13] | Self-consumption | yes | Over 20 | unknown |
| Xu et al. 2018 [11] | Energy trading | no | 1.1 to 10 | unknown |

Experimental validation of the findings of these studies has been lacking. For the revenue-maximising LM, profit-maximising LM and PBM respectively, linearly extrapolating the results from Fig. 5 suggests calendar lives of 1.4, 8.1 and 12 years, and cycle lives of 1980, 3150 and 7290 full equivalent cycles. These results fit well within those reported in Table 2 and confirm that simplistic control methods will in the worst case reduce lifetime to around 1 year while advanced degradation-aware control methods can extend it to over 10 years.

In Table 1, for the profit maximising linear model, the degradation prediction versus experiment has a relative error of 40% (i.e. difference between simulated and measured capacity at end of test, divided by the latter), whereas the physics-based model has a relative error of 19%. A large number of studies in the literature, especially the more economic- or application-oriented, use a model similar to the linear models validated here, and therefore their degradation predictions when compared with real batteries are likely to be quite inaccurate, implying that their lifetime predictions and conclusions may also be off the mark.

Turning to the economic modelling of storage, many studies have assessed the levelized cost or overall business case for stationary battery storage. A certain lifetime is assumed, and then the cost of using lithium-ion batteries to deliver a certain application is computed. Table 3 shows the assumed lifetimes. These are inputs to the economic models and they predominantly use a simple linear battery model. Comparing this to Table 2 reveals that these studies are optimistic compared to the results from optimisation studies. Our experimental results fall somewhere in the middle, with a best-case of 12 years and 7290 cycles with the cells tested. We suggest that this indicates that to achieve the performance assumed in economic studies without incurring additional costs (e.g. by oversizing systems), advanced models and control methods are essential.

Table 3: Overview of battery lifetime assumptions in economic studies

|  | Calendar life [year] | Cycle life [cycles] |
|---|---|---|
| Battke et al. 2013 [30] | 11.5 ± 1.5 | 10,250 ± 5,250 |
| Hiremath et al. 2015 [38] | unknown | 10,250 |
| Hittinger et al. 2012 [39] | 10 | unknown |
| Mostafa et al. 2020 [40] | 10 | 4,500 |
| Schmidt 2019. [41] | 13 years | 2,000 to 3,500 |
| Staffell et al. 2016 | 8 | 6,000 |
| Le Varlet et al. 2020 [42] | 15 to 21 | 5,840 to 30,000 |
| Zakeri et al. 2015 [43] | 10 to 20 | 4,500 to 6,000 |

## 4.2 Scale up from single cells to full battery systems

Due to computational and experimental limitations, many studies, including this one, focus on single cells, while real grid batteries consist of thousands of cells, plus additional equipment. This impacts assumptions about efficiency (and hence revenue) as well as degradation.

In terms of efficiency, the power-electronic converter greatly decreases the efficiency when the battery is operating at low power [44]. However, at the high power levels expected for wholesale arbitrage, the costs associated with converter losses are not expected to be significant relative to other costs [45]. Additional losses will be incurred in the thermal management system. This will however not significantly alter the results presented here since cooling is necessary for all cases.

In terms of degradation, although cell-to-cell variations of capacity and resistance exist, studies [46], [47] have shown that variability remains small until the battery has lost more than 20% capacity. Only after the so-called knee-point, which is usually after 80% remaining capacity, do cell-to-cell variations increase significantly. Existing cell-to-cell variations are also compensated for by the parallel connections present in nearly all battery systems, as well as by increasingly advanced balancing systems. Therefore, the overall conclusions of this study should hold for large-scale battery systems as well as for single cells.

## 5. Conclusions

The degradation and lifetime of lithium-ion batteries is heavily dependent on how they are used as a result of many interacting nonlinear physical processes. However, most techno-economic assessments of grid batteries use simplified battery models to facilitate design and operational optimisation. In this work, we bridged the gap between technical fidelity and economic dispatch in an energy trading application, investigating the economic effects of using physics-based battery models through experiments. The measurements confirm the improvements predicted in earlier work.

Conventional heuristic degradation modelling approaches limit the SoC operating window, reduce the power level, and decrease the energy throughput of the battery, which reduces degradation over time. However, a year-long experiment identified that using these conventional methods, degradation per cycle remains constant, implying that these methods cannot increase the number of cycles a battery can achieve over its lifetime. On the other hand, our experiments showed that a nonlinear physics-based model can decrease degradation both over time and per cycle by 30% respectively 70% compared to the heuristic approach.

At the same time as reducing degradation, the revenue using the physics-based approach increased by 20%. This indicates that there is no direct trade-off between increasing revenue and decreasing degradation, but that both can be achieved at the same time using more advanced models. Linearly extrapolating the results until the batteries have lost 20% of their capacity, the revenue over the entire lifetime from the physics-based approach is 70% higher than with the heuristic approach. This suggests that the economic potential for grid-connected lithium-ion batteries for energy trading might also be higher than estimated up to now. However, there are two challenges in scaling these approaches from the cell level tested here up to the larger system level and implementing them operationally. First the physics-based model approach is computationally more expensive than the linear model approach. Second, much closer integration between cell suppliers, system integrators and asset optimisers would be required, as well as tight coupling of the battery management system and energy management system.

Comparing the measured performance with the outcomes predicted by our earlier simulations shows that the model accuracy is dramatically improved using physics-based approaches, which reduced the prediction error for degradation from 1%-pts to 0.3%-pts, halving the relative error compared to simpler models. In terms of monetary value, this has a very significant effect: the error on the net profit was reduced from 170% to 13%. This shows how physics-based approaches can increase the confidence in upfront business assessments, reducing the risk associated with investment.

The increased accuracy of lifetime forecasts also demonstrates the predictive power of physics-based models for battery degradation and the importance of model-based control for unlocking value. The models used here were parametrised from standard lab degradation experiments but managed to accurately extrapolate ageing in a real-life usage scenario far from the standard test conditions. This stands in contrast with machine learning approaches, which in order to make accurate predictions would have to have been trained on a comprehensive ageing dataset measured under application-specific usage conditions.

## Supplementary materials

(1) Additional figures are given in the Supplementary Materials document.

(2) The dataset generated during this study is available at http://howey.eng.ox.ac.uk/data-and-code/. In due time, it will be deposited permanently at the Oxford Research Archive (https://ora.ox.ac.uk/).

## Acknowledgements

This work was supported by VITO. We also thank EIT InnoEnergy for providing the opportunity to participate in their PhD school. We are grateful to members of the BMWS community for providing helpful feedback and suggestions on this article.

## Author contributions

Conceptualisation, J.M.R., G.M, and D.A.H.; methodology, software, validation, and analysis, J.M.R; investigation, and resources, G.M.; Data curation, and writing – original draft, J.M.R; writing – review & editing, J.M.R, G.M. and D.A.H; visualisation, J.M.R; supervision, G.M, and D.A.H; project administration, J.M.R; funding acquisition, J.M.R, G.M, and D.A.H.

Appendix: Modelling and optimisation approach

This is a summary of the methods used in [15], repeated here to aid readability of this paper.

**Linear model**

The state of charge $z$ is the integral of the cell power over time $P(t)$, normalised by the nominal cell energy capacity $C_{\text{nom}}$ (10 Wh),

$$\frac{dz(t)}{dt} = \frac{P(t)}{C_{\text{nom}}}. \tag{A1}$$

The state of charge is in the range 0 to 1. Degradation reduces the actual capacity $C(t)$, and is accounted for by a linear function with fitting constants $\beta_1$ to penalise charge throughput and $\beta_2$ to penalise maximum power,

$$\frac{dC(t)}{dt} = \beta_1 |P(t)| + \beta_2 P^{\text{max}}, \tag{A2}$$

where $P^{\text{max}} = \max P(t)$ is an optimisation variable. The value of $\beta_1$ was set to $1.26 \cdot 10^{-5}$ h/s, equivalent to 8000 full equivalent cycles over the cell's lifetime, defined as the period until the cell has lost 20% of its capacity. The value of $\beta_2$ was set to $2.12 \cdot 10^{-4}$ h, which, for the power levels in this application, reduced the lifetime for wholesale arbitrage to about 2800 full equivalent cycles. Note that during the optimisation, $P^{\text{max}}$ was calculated over each optimisation horizon of two days separately, such that periods with large price variations over short time scales resulted in a higher value compared to periods with a smoother price profile.

**Physics-based model**

A summary of the single particle model is given here. In the equations below, the time dependency of state variables and derived values is omitted to simplify notation. Negative currents refer to charging, positive values indicate discharging. The terms negative electrode and anode are used interchangeably, as are the terms positive electrode and cathode. The main dynamic equation in the SPM relates to solid phase transport in spherical particles. The lithium concentration in electrode $i$ at radius $r$, $c_i(r)$, is calculated according to Fick's law of diffusion,

$$\frac{\partial c_i(r)}{\partial t} = \frac{D_i}{r^2} \frac{\partial}{\partial r}\left(r^2 \frac{\partial c_i(r)}{\partial r}\right), \tag{A3}$$

where $D_i$ is the diffusion constant of electrode $i$. Due to symmetry, the concentration gradient at the centre of the particle is zero. At the surface, radius $R_i$, the concentration gradient must be compatible with the lithium flux $j_i$, which is linked to the battery current $I$ through Faraday's constant $F$, the effective surface area $a_i$, the geometric surface area $A_i$ and thickness $\tau_i$,

$$\left.\frac{\partial c_i(r)}{\partial r}\right|_{r=R_i} = j_i = \frac{I}{F a_i A_i \tau_i}. \tag{A4}$$

The chemical intercalation reaction is described by the Butler-Volmer equation,

$$j_i = j_{i,0}\left(\exp\left(-\frac{\alpha F}{RT}\eta_i\right) - \exp\left(\frac{(1-\alpha)F}{RT}\eta_i\right)\right), \tag{A5}$$

where, $\alpha$ is the charge transfer coefficient, taken to be 0.5, $R$ is the ideal gas constant, $T$ is the cell temperature, $\eta_i$ is the overpotential, and $j_{i,0}$ is the exchange current density given by

$$j_{i,0} = nFk_i c_i(R_i)^\alpha c_e^{(1-\alpha)}\left(c_i^{\max} - c_i(R_i)\right)^{(1-\alpha)}. \tag{A6}$$

The concentration in the electrolyte $c_e$ and the maximum concentration in the electrode $c_i^{\max}$ are constant. The rate constant $k_i$ depends on the temperature according to an Arrhenius relationship,

$$k_i = k_i^{ref}\exp\left[\frac{E_{k,i}}{R}\left(\frac{1}{T} - \frac{1}{T^{\text{ref}}}\right)\right], \tag{A7}$$

as does the diffusion constant,

$$D_i = D_i^{ref}\exp\left[\frac{E_{D,i}}{R}\left(\frac{1}{T} - \frac{1}{T^{\text{ref}}}\right)\right], \tag{A8}$$

with the values at a reference temperature indicated by the superscript ref, and the activation energies by $E$.

A bulk thermal model similar to that developed by Guo et al. [17] was included,

$$\rho A\tau C_p\frac{dT}{dt} = I^2 R_{\text{tot}} + I(\eta_n - \eta_p) + IT\frac{\partial U}{\partial t} - hA(T - T_{\text{env}}), \tag{A9}$$

for a battery with density $\rho$, heat capacity $C_p$, total resistance $R_{\text{tot}}$, entropic coefficient $\frac{\partial U}{\partial t}$, convective coefficient $h$, and environmental temperature $T_{\text{env}}$.

The terminal voltage $V$ is a function of the open circuit potential of each electrode, $U_i$, which in turn is a function of the lithium concentration at the surface of each sphere, and the temperature. There is a voltage drop due to the chemical reactions at each electrode and the ohmic resistance of the cell,

$$V = U_p(c_p(R_p), T) - U_n(c_n(R_n), T) - (\eta_n - \eta_p) - IR_{tot}. \tag{A10}$$

A diffusion and kinetically limited model for SEI growth was included [19]. The side reaction current associated with the SEI layer growth has flux $j_{\text{sei}}$,

$$j_{\text{sei}} = \frac{1}{F}\frac{\beta_3 \exp\left(-\frac{\alpha_{\text{sei}}F}{RT}\eta_n\right)}{\dfrac{1}{Fk_{\text{sei}}\exp\left(-\frac{n_{\text{sei}}F}{RT}(U_n - 0.4)\right)} + \dfrac{\tau_{\text{sei}}}{FD_{\text{sei}}}}, \tag{A11}$$

and this is added to the boundary condition at the surface of the anode particle given by equation A4. The rate constant $k_{\text{sei}}$ and diffusion constant $D_{\text{sei}}$ are also temperature dependent, similar to equations A7 and A8. A fitting constant $\beta_3$ is added, although it could be incorporated in the rate and diffusion constants.

This side reaction removes cyclable lithium ions from the system, reducing the amount of charge that can be stored. Ignoring the nonlinearities in the voltage curve, the rate at which the energy capacity is lost is approximately the product of the nominal cell voltage (3.7 V) and the SEI current, which itself is the product of the SEI current density and the surface area as was the case in equation A4,

$$\frac{dC}{dt} = 3.7\, I_{sei} = 3.7\, j_{sei} F a_n A_n \tau_n \,. \tag{A12}$$

# Supplementary material

NOTE: The dataset associated with this article may be downloaded from Oxford Research Archive, DOI: 10.5287/bodleian:gJPdDzvP4 (URL https://doi.org/10.5287/bodleian:gJPdDzvP4 ).

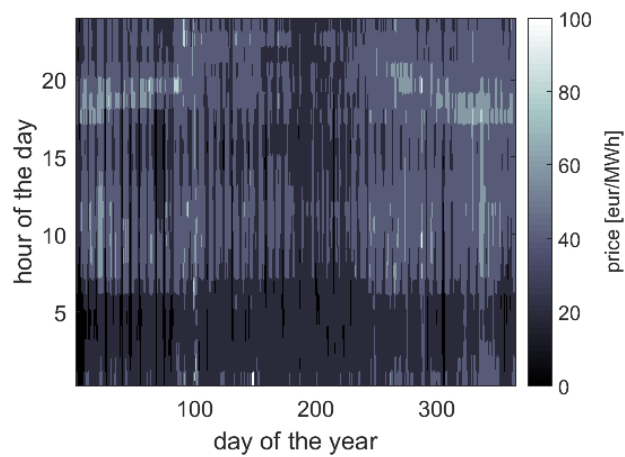

Figure S1: The price of electricity on the day-ahead market in Belgium in 2014.

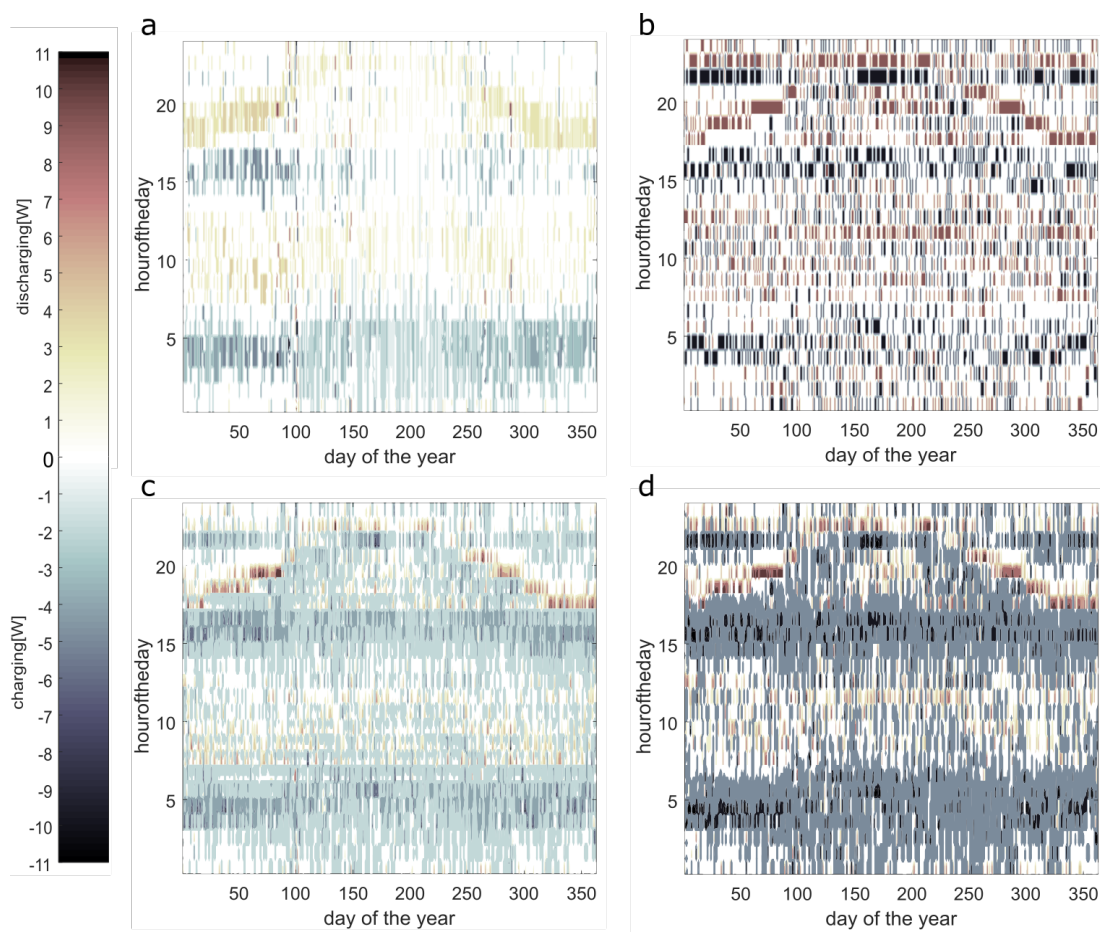

Figure S2: Usage profiles resulting from the four optimal solutions using: a linear model maximising profit; b linear model maximising revenue; c physics-based model maximising profit; d physics-based model maximising revenue.